There are two files. They are separated by three lines like the next one:
///////////////////////////////////////////////////////////////////////////
The first file is the body of the paper and the second one is the file of
style in LaTeX which has to be called espcrc2.sty on your directory.
\documentstyle[twoside,fleqn,espcrc2]{article}

\newcommand{\AmS}{{\protect\the\textfont2
  A\kern-.1667em\lower.5ex\hbox{M}\kern-.125emS}}

\title{Three-Loop Results on the Lattice}

\author{B. All\'es\address{Departamento de F\'{\i}sica Te\'orica y del Cosmos,
        Facultad de Ciencias \\
        Universidad de Granada, 18071-Granada, Spain}%
        \thanks{Presented the talk.}
        ,
        M. Campostrini\address{Dipartimento di Fisica dell'Universit\`a and
        I.N.F.N. \\
        Piazza Torricelli 2, 56126-Pisa, Italy}
        ,
        A. Feo$^{\rm b}$  and
        H. Panagopoulos$^{\rm b,}$\address{Department of Physics, University
        of Cyprus, Nicosia, Cyprus}}

\begin{document}

\begin{abstract}
We present some new three-loop results in lattice gauge theories, for the Free
Energy and for the Topological Susceptibility. These results are an outcome of
a scheme which we are developing (using a symbolic manipulation language), for
the analytic computation of renormalization functions on the lattice.
\end{abstract}

\maketitle

\section{INTRODUCTION}

The computation of vacuum expectation values of composite operators is
one of the most important tasks in lattice gauge theories. Examples of
such expectation values in QCD are the Condensates which are essential
tools for the SVZ sum rules \cite{SVZ79}
or the Topological Susceptibility which is
necessary to solve the so-called $U_A(1)$ problem and to understand the
$\eta'$ mass \cite{tH,W,V}.
In this talk we present some results related to the
computation on the lattice of the Gluon Condensate
\begin{equation}
G_2 \equiv \langle 0 | : {g^2 \over {4 \pi^2}} G_{\mu \nu}^{b}
G_{\mu \nu}^{b} : | 0  \rangle,
\end{equation}
and the Topological Susceptibility $\chi$
\begin{equation}
\chi \equiv \int \hbox{d}^4 x \langle 0 | \hbox{T}(Q(x) Q(0)) | 0 \rangle,
\end{equation}
where $Q(x)$ is the Topological Charge density
\begin{equation}
Q(x) \equiv {{g^2} \over {64 \pi^2}} \epsilon^{\mu \nu \rho \sigma}
G_{\mu \nu}^{b} G_{\rho \sigma}^{b}.
\end{equation}
In these equations, $g$ is the coupling constant of the QCD lagrangian and
$G_{\mu \nu}^{b}$ the strength tensor for the gluon fields.

To evaluate one of these quantities on the lattice, e. g.
$\langle 0 | {\cal A} | 0 \rangle$, one defines first a lattice
version ${\cal A}^L$ of ${\cal A}$ in such a way that
${\cal A}^L \enskip
{\buildrel {a \rightarrow 0} \over {\longrightarrow}} \enskip a^d {\cal A} +
{\cal O}(a^{d+1})$ where $d$ is the mass dimension of the operator ${\cal A}$
and $a$ is the lattice spacing.
For the Gluon Condensate and Topological
Susceptibility respectively, our choices for these lattice versions are
\begin{equation}
\begin{array}{l}
G_2^L \equiv \langle 0 | 1 - \Pi_{\mu \nu} | 0 \rangle,  \\
    \\
\chi^L \equiv \langle 0 | \sum_x Q^L(x) Q^L(0) | 0 \rangle,
\end{array}
\end{equation}
where $Q^L(x)$ is a lattice version of the Topological Charge density
\cite{DVF81}
\begin{equation}
\begin{array}{l}
    Q^{L}(x) = -{1 \over {2^4  32  \pi^2}}
    \sum_{\mu, \nu, \rho, \sigma = \pm 1}^{\pm 4}
    \epsilon_{\mu \nu \rho \sigma}  \times  \\
    \\
    \quad \quad \quad \quad
    Tr \left[ \Pi_{\mu \nu} (x)  \Pi_{\rho \sigma} (x) \right].
\end{array}
\end{equation}
In these equations, $\Pi_{\mu \nu}$ is the usual plaquette in the
$\mu - \nu$ plane.

Once the lattice version of the operator is defined, one can perform a
Monte Carlo simulation. The Monte Carlo data will give the physical
continuum value of $\langle 0 | {\cal A} | 0 \rangle$ modified by some
$a-$dependent renormalizations. These expressions for the Gluon Condensate
and Topological Susceptibility can be written in the following way
\cite{r8a,r8b,r9}
\begin{equation}
G_2^L = {{\pi^2} \over {12 N}} Z_G a^4 G_2 + \sum_{n \geq 1}
{{c_n} \over {\beta^n}} ,
\end{equation}
\begin{equation}
\chi^L = Z^2_Q a^4 \chi + a^4 G_2 \sum_{n \geq 2} {{b_n} \over {\beta^n}} +
\sum_{n \geq 3} {{d_n} \over {\beta^n}} ,
\end{equation}
where $N$ is the number of colors and as usual $\beta=2 N/g^2$. The lattice
spacing $a$ and $\beta$ are related by the renormalization group equation
\begin{equation}
a \Lambda_L = \left( {{2 N r_0} \over {\beta}} \right)^{ - r_1 /
 2 r_0^2} \exp\left( - {{\beta} \over {4 N r_0}} \right),
\end{equation}
where $\Lambda_L$ is the renormalization group invariant mass parameter
of QCD and $r_0$ and $r_1$ the first two coefficients of the $\beta$
function, $\beta(g) = -r_0 g^3 - r_1 g^5 - ...$

The last terms in these expressions (those proportional to $c_n$ and $d_n$)
are the perturbative tails and represent mixings with the unity operator.
The second term in Eq.(7) is a mixing with the Gluon Condensate. Finally,
$Z_G$ and $Z_Q$ are multiplicative finite renormalizations which relate
the lattice and the continuum definitions of the respective operators.
All of these coefficients can be computed in perturbation theory
and their knowledge is essential to extract the physical values $G_2$ and
$\chi$ from the Monte Carlo data and Eq.(6-7). The values of these
coefficients also depend on the lattice versions used for the operators and
the lattice action chosen. Throughout this work we have used the lattice
versions shown in Eq.(4) and the Wilson action.

\section{RENORMALIZATION CONSTANTS}

The first coefficients in the renormalization terms of
Eq.(6-7) can be calculated with rather small effort because they involve
very few Feynman diagrams. The result for $SU(3)$ is \cite{r8a,r8b,r9,r10,r11}
\begin{equation}
\begin{array}{l}
Z_G = 1 + {\cal O}(1/\beta^2), \\
Z_Q = 1 - 5.45/\beta + {\cal O}(1/\beta^2)  \\
b_2 = 6.32 \, \times \, 10^{-3}  \qquad d_3=3.58 \, \times \, 10^{-3}  \\
c_1 = 2.0 \qquad c_2=1.22 .
\end{array}
\end{equation}
However, the next coefficients involve many more diagrams. For instance,
$c_1$ and $c_2$ are calculated with three diagrams, but the next order,
$c_3$, needs the computation of 30 three-loop Feynman diagrams! To evaluate
these Feynman diagrams we have developed an algebraic computer program.
Major tasks in this algorithm are:

\noindent $i)$ Computing n-point vertices.

\noindent $ii)$ Producing a list of relevant diagrams, with the corresponding
weights.

\noindent $iii)$ Performing the contractions for each diagram, using up all
existing symmetries to produce a compact result.

\noindent $iv)$ Extracting the analytic dependence of each diagram on its
external momenta $p$, in the limit $ap \rightarrow 0$.

\noindent $v)$ Producing the optimized code for the numerical
calculation of the loop integrals.

Regarding the first step, some difficulties inherent to the lattice are:
The existence of vertices with an arbitrary number of gluons, a plethora
of ``tensor'' structures (due to lack of rotational invariance) and a great
proliferation in the size of vertices (a 6-point vertex in its most compact
form may typically require some dozens of output pages). This in turn
necessitates, in the third step, simplifying all intermediate expressions
as much as possible, by devising algorithms which use up the
symmetries of the diagram
under exchange of external legs, under allowed redefinitions of momenta and
under permutation of the (numerous) dummy indices.

The fourth step is necessary for computing multiplicative renormalizations.
The evaluation of these form factors is in progress.

\begin{table*}[hbt]
% space before first and after last column: 1.5pc
% space between columns: 3.0pc (twice the above)
\setlength{\tabcolsep}{1.5pc}
% -----------------------------------------------------
% adapted from TeX book, p. 241
\newlength{\digitwidth} \settowidth{\digitwidth}{\rm 0}
\catcode`?=\active \def?{\kern\digitwidth}
% -----------------------------------------------------
\caption{Exact and fitted values for $c_3$ and $d_4$}
\label{tab:effluents}
\begin{tabular}{lcccc}
\hline
                 & \multicolumn{2}{c}{$SU(2)$}
                 & \multicolumn{2}{c}{$SU(3)$} \\
\cline{2-3} \cline{4-5}
                 & \multicolumn{1}{c}{Exact}
                 & \multicolumn{1}{c}{Fitted}
                 & \multicolumn{1}{c}{Exact}
                 & \multicolumn{1}{c}{Fitted}         \\
\hline
$c_3$    & $ 0.16$ & $0.18$ & $  3.12$ & $  3.0$ \\
$d_4$    & $ 7.0 \, \times \, 10^{-5}$ & $1.8(6) \, \times \, 10^{-4}$ &
           $ 8.4 \, \times \, 10^{-4}$ & $1.6(1.0) \, \times \, 10^{-3}$ \\
\hline
\end{tabular}
\end{table*}

Now the final expression for the diagram can be integrated. With the code
produced by the algorithm, we can calculate the numerical value of the diagram
for rather small lattices and then extrapolate for larger lattices.
This extrapolation is performed by assuming the following dependence of the
diagram on the lattice size $L$
\begin{equation}
\hbox{diagram} = A + {{B} \over {L^n}}.
\end{equation}
The explicit values for $A$, $B$ as well as that of $n$ (which may vary for
different diagrams) are determined by the extrapolation. The result is then
confronted with the results for an infinite lattice, which we also compute.
We will explain this computation for an infinite lattice with an example.
Let us consider the following integral
\begin{equation}
I = \int^{+\pi}_{-\pi} {1 \over {\hat{p}^2}} {1 \over {\hat{q}^2}}
{1 \over {\hat{p+q}^2}} {{\hbox{d}^4p} \over {(2 \pi)^4}}
{{\hbox{d}^4q} \over {(2 \pi)^4}}
\end{equation}
which is the value of a two-loop diagram for an infinite lattice. In this
expression, $\hat{p}^2 = 2 \sum_{\mu} (1 - \cos p_{\mu})$.
Using the Schwinger representation for the propagators,
\begin{equation}
{1 \over {\hat{p}^2}} = \int^{\infty}_0 \hbox{e}^{- \alpha \hat{p}^2}
\hbox{d}\alpha \; ,
\end{equation}
we get
\begin{equation}
I = \int\limits^{\infty}_0 \hbox{d}\alpha_1
\int\limits^{\infty}_0 \hbox{d}\alpha_2
\int\limits^{\infty}_0 \hbox{d}\alpha_3  \; \;
\Phi^4(\alpha_1, \alpha_2, \alpha_3)
\end{equation}
where $\Phi$ is
\begin{equation}
\begin{array}{l}
\Phi(\alpha_1, \alpha_2, \alpha_3) = \\
   \\
\hbox{e}^{-2(\alpha_1+\alpha_2+\alpha_3)}
\int^{+\pi}_{-\pi} {{\hbox{d}p} \over {2 \pi}}
\int^{+\pi}_{-\pi} {{\hbox{d}q} \over {2 \pi}}  \times  \\
   \\
\exp(2 \alpha_1 \cos p + 2 \alpha_2 \cos q + 2 \alpha_3 \cos (p+q)) = \\
   \\
\hbox{e}^{-2(\alpha_1 + \alpha_2 + \alpha_3)} \int^{+\pi}_{-\pi}
{{\hbox{d}p} \over {2 \pi}}  \times \\
   \\
\hbox{e}^{2 \alpha_2 \cos p} \; \; \;
\hbox{I}_0(2 \sqrt{\alpha_1^2 + \alpha_3^2 +
2 \alpha_1 \alpha_3 \cos p} ).
\end{array}
\end{equation}
In Eq.(14) $\hbox{I}_0$ is the modified Bessel function.
Therefore we have reduced the number of integration variables from 8
in Eq.(11) to 4 in Eq.(13). For three-loop integrals with 12 integration
variables, this procedure reduces this number to 6 integrations (for a few
cases it only reduces to 8 integrations).
Now, a numerical integration is feasible.

The extrapolation done with Eq.(10) is enough to obtain the correct values
of the coefficients with three digits.
Further details of the algorithm will be published elsewhere.

\section{RESULTS}

Summing up the individual contributions of all diagrams we obtain for
the gauge group $SU(N)$ and an infinite lattice \cite{r12}
\begin{equation}
d_4 = N^4 (N^2 - 1) (1.73 N^2 - 10.83 + {{73.83} \over {N^2}} ) 10^{-7},
\end{equation}
\begin{equation}
c_3 = N^2 (N^2 - 1) (7.28 N^2 - 27.35 + {{46.15} \over {N^2}} ) 10^{-3}.
\end{equation}
The exact values for the coefficients obtained with our scheme
can be compared with the values extracted by a best fit of all Monte Carlo
data with Eqs.(6-9). Both values are shown in table 1 for the
coefficients $d_4$ and $c_3$ and for an infinite lattice \cite{r8a,r8b,r9,r15}.

The agreement between fitted and exact results is manifest for $c_3$.
The data for the Topological Susceptibility had a rather low statistics,
therefore and within the errors the numbers shown in table 1 for $d_4$
are in acceptable agreement.

Finally, we can add the exact values of $c_3$ and $d_4$ to those of
Eq.(9) and perform again the best fits. The values extracted for $G_2$
and $\chi$ are in complete agreement with those reported in refs.
\cite{r8a,r8b,r9,r15}
\footnote[1]{Actually, for the Topological Susceptibility
we calculate the unrenormalized value $\chi_u$
because the physical one is better obtained by
using the cooled data of ref.\cite{r15}.}.

\section{ACKNOWLEDGMENTS}

We wish to thank Adriano Di Giacomo for useful conversations and the
spanish-italian ``Acci\'on Integrada/Azione Integrata'' number $A17$ for
financial support. B. A. also acknowledges a spanish CICYT contract.

\end{document}